\documentstyle[sprocl]{article}
 
\def\be{\begin{equation}}
\def\ee{\end{equation}}
\def\bea{\begin{eqnarray}}
\def\eea{\end{eqnarray}}
\def\pf{\partial_\varphi}
\begin{document}
\begin{flushright}
YITP--98--47\\
July 31, 1998\\
\end{flushright}
\vskip2cm
\title{EINSTEIN-YANG-MILLS SOLITONS: TOWARDS NEW DEGREES OF FREEDOM
\footnote{Extended version of a talk given at the International Workshop
``Mathematical cosmology'', Potsdam,   March 30 -- April 4, 1998.}}

\author{D. V. GAL'TSOV \footnote{email: gdv@yukawa.kyoto-u.ac.jp,
galtsov@grg.phys.msu.su}}

\address{Yukawa Institute for Theoretical Physics,\\
Kyoto University, Kyoto 606, Japan\\ and}
\address{Department of Theoretical Physics,
       Moscow State University,\\ 119899 Moscow, Russia}

\maketitle\abstracts{A recent progress in obtaining non-spherical
and non-static solitons in the four-dimensional 
Einstein--Yang--Mills (EYM) theory is
discussed, and a non-perturbative formulation of the stationary
axisymmetric problem is attempted. First a 2D dilaton gravity 
model is derived for the spherically symmetric time-dependent 
configurations. Then a similar Euclidean representation is  constructed 
for the stationary axisymmetric non-circular SU(2) EYM system using the 
(2+1)+1 reduction scheme suggested by Maeda, Sasaki, Nakamura and Miyama.
The crucial role in this reduction is played by the extra terms
entering the reduced Yang--Mills and Kaluza--Klein two-forms
similarly to Chern--Simons terms in the theories with higher rank
antisymmetric tensor fields. We also derive a simple
2D action describing static axisymmetric magnetic EYM configurations
and discuss a possibility of existence of cylindrical EYM sphalerons.}

\section{Introduction}

Investigation of classical solutions to 4D gravity coupled
non-Abelian gauge theories, inspired by the discovery of particle-like
solutions to the Einstein-Yang-Mills (EYM) equations by Bartnik and
McKinnon~\cite{bk}, turned out to be a fruitful field of research
which revealed many unexpected features both in the black hole and soliton
physics. EYM black holes~\cite{vgkb}  gave rise to  critical revision
of some conceptual foundations of the black hole theory, especially of
such folkloric beliefs as the no-hair and uniqueness conjectures.
A more recent progress in this area was due to investigation of 
particle-like and black hole solutions
without spherical symmetry. It was shown that the celebrated Israel's
theorems, implying that static vacuum and electrovacuum black holes
are spherically symmetric~\cite{isr}, are no longer true for non-Abelian
gravity coupled theories. Namely, in the EYM theory the static solitons 
and black holes with higher winding numbers are only axially 
symmetric~\cite{klku}, like multimonopole and multisphaleron  solutions
in the flat space. In more sophisticated models even axial symmetry 
no longer holds for static black holes~\cite{rdw}.

A substantial progress was also achieved recently in understanding the nature 
of {\em rotating} solutions. It was shown by Volkov and Straumann~\cite{vs}
that electrically charged SU(2) black holes, prohibited in the static 
spherically symmetric case by a `non-abelian
baldness' theorem~\cite{ge}, still may
exist, but they are {\em non-static} and only axially symmetric. Stationary
EYM {\em black holes} are likely to increase the number of characterizing
them independent parameters (a mass and a node number in the static spherically
symmetric case) by two: an angular momentum and an electric charge. In contrary, 
for the {\em regular} rotating EYM sphalerons these two parameters were found 
to be related~\cite{bhsv}, thus the electric charge has to be
regarded rather as an induced one.
It was also claimed that in the theories including Higgs fields no rotating
particle-like solutions may exist at all~\cite{BH,HSV}, perhaps unless
the angular momentum takes discrete values. The corresponding rotating black
holes necessarily acquire electric charges.

These intriguing results for the stationary axisymmetric non-Abelian
self-gravitating solutions were
obtained, however, only perturbatively. In the existing literature no fully 
non-linear analysis is available for the stationary axisymmetric EYM system 
(apart from some preliminary considerations in~\cite{BH}),
and even the corresponding ansatz for the YM field
is not exhibited yet. The essential complication in the non-Abelian
case with respect to various Abelian theories is that now the {\em circularity}
assumption for the spacetime metric should be avoided, as was argued by
Heusler and Straumann~\cite{HS}. Therefore a general analysis of the EYM
system for stationary axisymmetric non-circular spacetime is necessary for 
further progress. Here we  present such an analysis in terms of the Kaluza-Klein 
reduction, which ends up in a derivation of an Euclidean 2D dilaton gravity 
model coupled to non-Abelian scalar and vector fields inherited from the 4D
SU(2) YM field, as well as to Kaluza-Klein two-forms and scalar
moduli. We also derive a simple representation for the static
axially symmetric magnetic configurations and discuss one curious
cylindrically symmetric truncation.

\section{Spherical EYM system as 2D dilaton gravity}
Throughout the paper we will be dealing with the 4D SU(2)
Einstein-Yang-Mills (EYM) system described by the action
(up to a boundary term)
\be \label{EYM}
S_{\rm EYM}=(16\pi)^{-1} \int \left\{  R_4 +
2 {\rm tr}\, F_{\mu\nu}F^{\mu\nu}\right\}
\sqrt{-g_4}\, d^4 x\, ,
\ee
where an antihermitean representation of the algebra is understood,
and the gravitational and gauge constants are removed by an appropriate
coordinate rescaling. 
As a warming up exercise, let us consider the dimensional reduction of 
(\ref{EYM})  for the spherically symmetric time-dependent spacetimes.
We will adopt the following parameterization of the spacetime metric
$ds^2=g_{4\mu\nu}dx^\mu dx^\nu$ :
\be \label{2+2}
ds^2=g_{ab}dx^a dx^b + {\rm e}^{-2\psi}
(d\vartheta^2 + \sin^2\vartheta d\varphi^2),
\ee
where $g_{ab},\; \psi$ are functions of $x^a=t,r$.
The Yang--Mills potential one-form can be decomposed correspondingly
as $A=a+\Phi$, where
\be \label{6}
a=a_b \,\tau_r dx^b
\ee
is a `dynamical' part, parameterized by a 2D  real Abelian one-form $a_b$
depending on $x^a$, and
\be   \label{7}
\Phi={\rm Re} \left\{
(w-1)(\tau_\varphi -i\tau_\vartheta)
(d\vartheta-i\sin\vartheta d\varphi)\right\}
\ee
is an effective Higgs field parameterized by  a complex-valued
function $w(x^a)$. We use the rotated basis
\be
\tau_r=
\sin\vartheta(\cos\varphi\, \tau_1+
 \sin\varphi\, \tau_2)+
\cos\vartheta\, \tau_3, \quad
\tau_{\vartheta}=\partial_{\vartheta}\tau_r, \quad
\tau_{\varphi}=
\frac{1}{\sin\vartheta}\, \partial_{\varphi}\tau_r,
\ee
and $\tau_k=\sigma_k/(2i),\, k=1,2,3$, with $\sigma_k$ being the Pauli matrices.

As it is well-known, the dimensional reduction of the Einstein-Hilbert action
under an assumption of spherical symmetry leads to the 2D
dilaton gravity action
\be \label{JT}
S_G=\frac14\int \left\{e^{-2\psi}\left( R
+ 2(\partial \psi)^2 \right)+ 2 \right\}\sqrt{|g|}\,d^2x,
\ee
where $d^2x=drdt$ and $R$ is the  2D scalar curvature. 
For the reduced YM field one gets the
dilaton coupled version of the 2D scalar electrodynamics, in which $w$ enters
as a complex scalar, while  $a_b$ --- as  a  U(1) gauge field:
\be \label{S2YM}
S_{YM}=-\int\left\{\frac14 e^{-2\psi}f_{ab}  f^{ab}+ |{\hat D}w|^2+\frac12
e^{2\psi}(1-|w|^2)^2\right\}\sqrt{|g|}\, d^2x,
\ee
where
\be
f_{ab}=\partial_a a_b -\partial_b
 a_a,\quad {\hat D}_b w=\partial_b  w-ia_b w.
\ee

The sum of the actions  $S_G+S_{YM}$ describes the dynamics of the moving
self-gravitating   spherical shells of the YM field. The system
is thus equivalent to the 2D Abelian Higgs model coupled to a
Jackiw-Teitelboim type theory. Note that  no cosmological term is present
in (\ref{JT}), instead there is a constant term not multiplied by the dilaton.
The action is
invariant under the 2D diffeomorphisms and has a U(1) gauge symmetry
\be \label{U1}
w \rightarrow e^{ i \alpha} w, \quad
a_b\rightarrow a_b +\partial_b  \alpha,
\ee
where $\alpha(x^a)$ is a gauge parameter. In 2D dilaton gravity it is usual
to use a diffeomorphism invariance in order to impose a conformal gauge
$g_{ab}=\exp(2\rho)\eta_{ab}$,
but  here we prefer to
leave the metric coefficients unspecified (for reduction
of the spherical EYM system in a different fixed gauge see~\cite{BHS}).

The total energy-momentum tensor including the dilaton part is
\be
T_{ab}={\rm e}^{-2\psi} \left(\partial_a \psi \;\partial_b \psi-
\psi_{;ab}\right)+T_{ab}^{YM} \;\; + {\rm trace \;terms},
\ee
where $T_{ab}^{YM}$ corresponds to a variation of (\ref{S2YM}).
Since the 2D Einstein tensor vanishes, 2D Einstein's equations
take the form
\be
T_{ab}=0,
\ee
encoding the $(t,r)$ components of the 4D Einstein's equations, the remaining
part of which is contained in the equation for the `dilaton'.
The matter equations can be obtained by a variation of the action,
after that one can fix the YM and gravitational gauge using the 
three-parameter gauge group described above.

\section{(2+1)+1 reduction for non-circular spacetime}
Our purpose is to derive a two-dimensional representation for
the {\em stationary axisymmetric} EYM system. The corresponding spacetime
admits two commuting Killing vectors, which may be written in appropriate 
coordinates as
\be
K=\partial_t, \quad \tilde K =\partial_\varphi.
\ee
For most of known explicitly stationary
axisymmetric solutions in general relativity these  Killing vectors
satisfy Frobenius conditions
\be \label{fro}
K_{[\mu}\tilde K_\nu \,\tilde K_{\lambda ]; \tau}=0,\quad
\tilde K_{[\mu}K_\nu \,K_{\lambda ] ; \tau}=0,
\ee
which mean that they are {\em hypersurface orthogonal}.
In this case the metric admits a particularly simple Lewis-Papapetrou
parameterization by three real-valued functions of two coordinates.

The {\em circularity theorem}~\cite{PA,CA} asserts that the necessary
and sufficient conditions
for Eqs.~(\ref{fro}) to hold are provided by the {\em Ricci-circularity} conditions
(for a more recent discussion see~\cite{hsbook}):
\be
K_{[\mu} \tilde K_\nu \,R_{\lambda]}^\tau \,\tilde K_\tau=0,\quad
\tilde K_{[\mu}  K_\nu \,R_{\lambda]}^\tau \, K_\tau=0.\quad
\ee
Since the Ricci tensor $R_{\lambda}^\tau$ is involved here, the validity
of (\ref{fro}) depends strongly on the matter system. In particular,
these conditions are satisfied in  the Einstein--Maxwell  theory  provided
the Maxwell field is also stationary and axially symmetric. For the EYM
system, as was discussed by Heusler and Straumann~\cite{HS}, this is
not the case, however. In order to fulfill the Ricci conditions, 
the YM field strength should have the components 
$F_{t\varphi}, \;\tilde F_{t\varphi}$  vanishing, 
but this is not true already for 
a spherically symmetric configuration (\ref{6}-\ref{7}).
Therefore, a sufficiently general EYM system with two commuting
Killing vectors can not be described  by the Lewis-Papapetrou ansatz.

Here we show that an appropriate dimensional reduction of the stationary
axisymmetric non-circular EYM system leads to some 2D dilaton gravity
model which, though being more complicated
than that in the spherical case, still looks tractable.
We will follow the (2+1)+1 approach suggested by Maeda, Sasaki, Nakamura
and Miyama~\cite{MSNM}  consisting in a two-step dimensional reduction
of the metric with respect to two commuting Killing vector fields. 
In their case the first
reduction was performed with respect to the spacelike Killing field, what
was suitable for treatment of an axisymmetric gravitational collapse.
An alternative reduction, first in the timelike direction, was employed   
by Gourgoulhon and Bonazzola \cite{GB} aiming to attack the problem 
of a non-circular magnetized star.
In both cases the dimensional reduction was effected at the level of
Einstein {\em equations}. The resulting systems are still too complicated
to be used with the YM matter source. We will see that a considerable
simplification may be achieved if one performs dimensional reduction
at the level of the {\em action}. An additional advantage of  our procedure
is that no gauge fixing is necessary at  any stage of the computation, so
one is free to use different gauge choices to further simplify the final
system.

As the first step let us perform  reduction with respect to $K$, representing
the 4D metric ($x^\mu=t, x^i$) as
\be
ds^2=-{\rm e}^\psi\left(dt+v_i dx^i\right)^2+{\rm e}^{-\psi} dl_3^2,
\quad dl_3^2=h_{ij}dx^i dx^j.
\ee
This reduction  for the EYM and EYMH systems was described  recently by
Brodbeck and Heusler~\cite{BH}. The decomposition of the Einstein term is
standard:
\be \label{R43}
\sqrt{-g_4}R_4=\sqrt{h}\left(R_3-\frac12 (\partial \psi)^2+
\frac12{\rm e}^{2\psi} \Omega_{ij}\Omega^{ij}\right)\;\;+\;\; {\it div},
\ee
where $R_3$ is the three--dimensional Ricci scalar, 
'${\it div}$' stands for a total divergence term, and
 $\Omega_{ij}$ is the Kaluza-Klein  two-form
\be \label{Om}
\Omega_{ij}=\partial_i v_j - \partial_j v_i.
\ee

To reduce the  material part it is convenient to use for the matrix-valued YM
potential  a standard Kaluza-Klein  split
\be \label{A1}
A=A_\mu dx^\mu=\Phi \left(dt+v_i dx^i\right) +{\cal A}_i dx^i,
\ee
where ${\cal A}$ is a purely three--dimensional gauge field one-form, 
while $\Phi$ plays a role of a three--dimensional Higgs field. It could be 
expected that, once the potential is split with respect to some Killing 
symmetry, one should demand an independence of its components on the 
coordinates along the Killing orbits, as it is usual for toroidal reduction 
in the Abelian  case. For a
non-Abelian field, however, this might be  in general too restrictive. In
fact, the symmetry of the gauge field under a spacetime isometry means
that the action of an isometry can be compensated by a suitable gauge
transformation, what  at the infinitesimal  level reads
\be \label{sym}
{\cal L}_K \, A_{\mu}=D_{\mu}W_{m},
\ee
with  $W$ being a Lie-algebra valued
gauge function. In the case of a unique timelike Killing vector one
can always choose $W=0$, and hence  $\cal A$ and $\Phi$ in (\ref{A1}) may be 
regarded as functions  of $x^i$ only. But for the second
Killing vector $\tilde K$ we will prefer to avoid such a choice and
allow for more general gauges.

Evaluating the field strength  corresponding to (\ref{A1})
one finds
\be \label{F43}
F=dA+A\wedge A={\cal F} + {\cal D} \Phi\wedge (dt+v_idx^i),
\ee
where the first term represents the three-dimensional components of the
full tensor
\be \label{cF}
{\cal F}_{ij} =\partial_i {\cal A}_j -\partial_j {\cal A}_i
+\left[{\cal A}_i, {\cal A}_j\right] + \Phi \Omega_{ij},
\ee
 while  the second --- mixed components, which take the form of
a 3D covariant derivative of the scalar $\Phi$
with respect to the connection ${\cal A}$:
\be
{\cal D}_i \Phi=\partial_i\Phi+\left[{\cal A}_i, \Phi\right].
\ee
An effective 3D field strength (\ref{cF})
contains an extra term proportional to the Kaluza-Klein  two-form
(\ref{Om}), which arises similarly to Chern-Simons terms
in the  non-diagonal reduction of higher rank antisymmetric forms
in Kaluza-Klein supergravities.

Now it is simple to find the following relation between  four and
three--dimensional YM lagrangians:
\be
\sqrt{-g_4}\; {\rm tr} \;F_{\mu\nu}F^{\mu\nu} =
\sqrt{h}\; {\rm tr}\;\left({\rm e}^\psi {\cal F}_{ij}{\cal F}^{ij}
-{\rm e}^{-\psi} \;{\cal D}_i \Phi {\cal D}^i \Phi\right),
\ee
where it is understood that all 4D indices are raised by
the  4D metric, while the 3D ones --- by the inverse to $h_{ij}$.
Therefore the resulting 3D EYM action reads~\cite{BH} (an overall 
coefficient being omitted):
\be
S=\int\left\{R_3-\frac12 (\partial \psi)^2+
\frac12{\rm e}^{2\psi} \Omega_{ij}\Omega^{ij}
+2 {\rm tr}\left({\rm e}^\psi {\cal F}_{ij}{\cal F}^{ij}
-{\rm e}^{-\psi} \;{\cal D}_i \Phi {\cal D}^i \Phi\right)
\right\}\sqrt{h}\, d^3x.
\ee
It describes a Euclidean system of 3D Yang-Mills and  Higgs fields (without
potential), a dilaton and an Abelian two-form coupled to gravity.
The equation for the Kaluza-Klein field strength following from this action 
assumes the form of the 3D Maxwell equations
\be
\partial_i\left({\rm e}^{2\psi}\sqrt{h}\;\Omega^{ij} +
4 {\rm e}^{\psi}\sqrt{h}\; {\rm tr}\;({\cal F}^{ij} \, \Phi)\right)=0,
\ee
which can be solved by introducing the twist potential~\cite{BH}
\be
\Omega^{ij} +
4 {\rm e}^{-\psi} \; {\rm tr}\;({\cal F}^{ij} \, \Phi)=
\frac{{\rm e}^{-2\psi}}{\sqrt{h}}\epsilon^{ijk}\partial_k \chi.
\ee
We will not, however, pursue this direction further, but rather pass
to the second step of reduction in terms of the Kaluza-Klein two-form 
$\Omega_{ij}$ itself.

By virtue of an {\em axial} symmetry, the 3D  metric can now be parameterized
as follows:
\be
dl_3^2={\rm e}^{2\phi}\left(d\varphi+k_a dx^a\right)^2+g_{ab} dx^a dx^b,
\ee
where a new two--dimensional Kaluza-Klein one-form $k_a$ and the 2D metric
$g_{ab}$ depend only on $x^a,\; a=1,2$. The one-form $k_a$ generates a field
strength
\be
\kappa_{ab}=\partial_a k_b - \partial_b k_a,
\ee
which is actually a scalar, $\kappa_{ab}=\kappa \epsilon_{ab}$.
The 3D  curvature scalar decomposes as
\be
\sqrt{h}  R_3=\sqrt{g}{\rm e}^\phi \left(R  -\frac12 \kappa_{ab}\kappa^{ab}
{\rm e}^{2\phi}\right) + {\it div},
\ee
where $R$ is the 2D Ricci scalar.
The action has the form of the Euclidean Jackiw-Teitelboim action
without cosmological constant. In order to reduce the  $\Omega_{ij}^2$
term in (\ref{R43}), first one has  to decompose the Kaluza-Klein one-form
\be
v_i dx^i= \omega \left(d\varphi + k_a dx^a\right) + \nu_a dx^a,
\ee
where a 2D scalar $\omega$ and a one-form $\nu_a$
are $\varphi$-independent.
In the reduced action the  2D field strength corresponding to $\nu_a$
acquires an extra term proportional to $\kappa_{ab}$:
\be
\omega_{ab}=\partial_a \nu_b - \partial_b \nu_a + \omega \kappa_{ab},
\ee
 so that  
\be
\sqrt{h}\Omega_{ij} \Omega^{ij}=\sqrt{g}\left({\rm e}^{-\phi}\partial_a \omega
\partial^a \omega  +  {\rm e}^\phi \omega_{ab} \omega^{ab}\right).
\ee

To further reduce the YM part of the 3D action one has to split the
matrix-valued YM one-form in a way similar to (\ref{A1}):
\be
{\cal A}= \Psi \left(d\varphi+k_a dx^a\right) + a_b dx^b.
\ee
Now we will no more assume the 2D matrix-valued scalar $\Psi$ and the
one-form $a_b$ to be $\varphi$-independent, but rather denote
\be
\pf \Psi=\Psi',\quad \pf a_b=a_b',
\ee
these derivatives will enter the  final formulas explicitly.
Similarly to (\ref{F43}), we obtain the following decomposition of the
three--dimensional field strength
\be
{\cal F}=f +\left(D'_a \Psi +\partial_a \omega \Phi-a'_a\right)dx^a
\wedge (d\varphi+k_a dx^a),
\ee
where an effective 2D field strength is
\be \label{F32}
f_{ab}=\partial_a  a_b-\partial_b a_a + [a_a,a_b]+ a'_a k_b-a'_b k_a+
\Phi \omega_{ab} +\Psi\kappa_{ab}.
\ee
A primed covariant derivative is defined as follows:
\be \label{pr}
D_b'\Psi=D_b\Psi - k_b \Psi' ,
\ee
the first term being the usual 2D covariant derivative
with respect to the connection~$a_b$
\be
D_b=\partial_b+\left[a_b,\;\;\right].
\ee
An extended field strength (\ref{F32}) now contains  extra terms proportional
to the Kaluza-Klein two-forms and also gauge-dependent terms whose emergence is
due to an allowed additional dependence of the gauge potential on $\varphi$.
This gauge dependence, however, should not cause any trouble once
the final stationary axisymmetric ansatz is fixed, but it will just give
more flexibility in the choice of a gauge  for this ansatz (the final
choice is outside the scope of the present paper).

Finally, the 3D gauge covariant derivative ${\cal D}_i \Phi$ splits
into the corresponding 2D derivative (primed)
and a commutator term $[\Phi,\Psi]$ combined with $\Phi'$.
Collecting all terms one obtains the following 2D dilaton gravity model:
\be \label{S2}
S_2=\int{\rm e}^\phi \left(R + L_m\right)\sqrt{g}\, d^2x,
\ee
with $L_m$ accounting both for YM and Kaluza-Klein variables:
\begin{eqnarray}\label{voila}
L_m&=&2{\rm tr}\Bigl\{\left[
f_{ab}f^{ab}+{\rm e}^{-2\phi}(D'_a\Psi+ \Phi\partial_a \omega  -a'_a)\, 
(D'^a\Psi+ \Phi\partial^a \omega -a'^a)\right]
{\rm e}^\psi \nonumber\\
&&\quad -\left[D'_a\Phi\, D'^a\Phi +{\rm e}^{-2\phi}
(\Phi' + [\Phi,\Psi])^2\right]{\rm e}^{-\psi}\Bigr\} \nonumber\\
&+& \frac12 \left({\rm e}^{2\psi}\omega_{ab}\omega^{ab} -
{\rm e}^{2\phi}\kappa_{ab}\kappa^{ab} +
{\rm e}^{2(\psi-\phi)}\partial_a \omega\, \partial^a \omega -
\partial_a \psi\, \partial^a\psi
\right),
\end{eqnarray}
a primed covariant derivative of $\Phi$ being  defined 
in the same way as (\ref{pr}).

Thus the stationary axially symmetric non-circular 4D EYM system 
gives rise to an Euclidean 2D dilaton gravity model, in which the matter 
(apart form the `dilaton' $\phi$) includes  the 2D Yang-Mills field, 
two Higgs fields (now with a quatric interaction term), 
two scalar moduli $\psi,\omega$, and two   Kaluza-Klein 
two-forms $\omega_{ab},\kappa_{ab}$.
(Note that, although no derivatives of $\phi$  appear explicitly, they
will be produced after elimination of second derivatives
entering the scalar curvature.)
Again, by virtue of the 2D Einstein equations,
the corresponding 2D energy-momentum tensor should vanish,
$T_{ab}=0$, the rest of the field equations being obtainable by variations
of the action over the `matter' fields. Purely gravitational 
degrees of freedom (the third line in (\ref{voila})) include three
scalars, two 2D two-forms (one degree of freedom each) and a 2D metric
which remains unspecified. It is instructive to write down the entire
4D metric in terms of the variables introduced:
\be
ds^2=-{\rm e}^\psi \left[dt+\omega d\varphi +(\nu_a+\omega k_a)dx^a\right]^2
+ {\rm e}^{2\phi-\psi}
(d\varphi+k_adx^a)^2+{\rm e}^{-\psi}g_{ab}dx^adx^b.
\ee
The decomposition of the full 4D YM connection reads:
\be
A=\Phi \left[dt+\omega d\varphi +(\nu_a+\omega k_a)dx^a\right]+
 \Psi (d\varphi+k_adx^a) + a_b dx^b.
\ee

\section{Static case: axial and cylindrical symmetry}

A specification of the stationary axisymmetric YM ansatz requires a study of
transformations of the matrix-valued quantities under spacetime isometries.
This will be presented elsewhere, while here we will concentrate
on a simpler case of the {\em static} configurations. Let us construct a 
2D dilaton gravity model for a purely magnetic 4D ansatz~\cite{RR}
used by Kleihaus and Kunz~\cite{klku} in the numerical investigations of
static solutions with multiple winding numbers $\nu$. In our notation
this corresponds to $\Phi=0$ and
\be \label{rr}
a_b=a_b \tau_\varphi,\quad
\Psi={\rm Re}\;w\, \tau_\rho+ \left({\rm Im}\;w-\nu\right)\tau_z,
\ee
where $w$ is a complex-valued function of $\rho,\,z$,  
\be\label{taunu}
\tau_\rho=\tau_x \cos \nu\varphi +\tau_y \sin \nu\varphi,\quad
\tau_\varphi=-\tau_x \sin \nu\varphi +\tau_y \cos \nu\varphi,
\ee
and we have denoted an Abelian one-form $a_b=a_b(\rho,\,z)$ by the same symbol.
The ansatz (\ref{rr}) has a U(1) symmetry (\ref{U1}), now with 
$\alpha=\alpha(\rho,\,z)$,
suggesting an Abelian covariant derivative ${\hat D}_b=\partial_b-ia_b$.

In the static case one has $k_a=\nu_a=\omega=0$  implying
 $\omega_{ab}=0=\kappa_{ab}$, and the 2D dilaton gravity action simplifies
dramatically:
\be \label{Sax}
S=\int\left\{{\rm e}^\phi\left(R-\frac12 \partial_a\psi \,\partial^a\psi\right)
+{\rm e}^{(\psi-\phi)}{\hat D_a w}{\bar{\hat D^a}} \bar w+
{\rm e}^{(\psi+\phi)}f_{ab}f^{ab}\right\}\sqrt{g}d^2x,
\ee
where $f_{ab}=\partial_a a_b-\partial_b a_a$. (Note that for an ansatz
(\ref{rr}) $a'_b$ and $\Psi'$ are non-zero.)
 Contrary to the spherical case, now there is no self-interaction
term for $w$, so we deal with  the {\em linear} scalar electrodynamics
in Euclidean space interacting with three extra scalar fields.

A variation over the metric gives Einstein equations $T_{ab}=0$, where 
the total energy momentum tensor can be read off from (\ref{Sax}).
The remaining matter equations (including those for the scalar fields) 
may then be written in the conformal gauge
\be \label{cfmet}
g_{ab}dx^a dx^b={\rm e}^\eta (d\rho^2+ dz^2),
\ee
Instead of writing down the equations, we just give the matter action 
in this fixed gauge, in which it reduces to a simple flat space action  
\be \label{Sm}
S_m=\int\left\{{\rm e}^{\phi} \partial\phi\cdot\partial\eta -
 \frac12 (\partial\psi)^2 +{\rm e}^{(\psi-\phi)}|{\hat D w}|^2 +
{\rm e}^{(\psi+\phi-\eta)}f_{ab}^2 \right\} \,d^2x,
\ee
where now all contractions correspond to the metric $\delta_{ab}$.
Note that  the total number of the four-dimensional gravitational degrees 
of freedom is three ($\psi,\,\phi,\, \eta$). The total number of 
the YM degrees of freedom is also three: four real functions 
minus one gauge.   
 
A similarity with  the scalar electrodynamics suggests  the existence of
a {\em cylindrical} YM ansatz which could simulate
the Nielsen-Olesen vortex. This is indeed the case: just choose
the gauge Re$\;w$=0 and set $a_\rho=0$, the remaining functions
$a_z=R(\rho),\;$ and Im$\;w=P(\rho)$ will do the job. If one omits
all gravitational variables, the equations of motion in the flat space will
exactly coincide with the Nielsen-Olesen equations, but now with the
self-interaction term switched off. Recall that the BPS limit
for an Abelian vortex corresponds to a non-zero value of the self-interaction
constant. For zero coupling no finite energy solutions exist, indeed, we
are dealing with the flat space Yang--Mills theory, in which
classical glueballs are prohibited. But still one can hope that gravity
will glue the vortex as it does for the spherical EYM sphaleron,
perhaps at the expence of loosing an asymptotic flatness.
Note that our cylindrical ansatz is not strictly two--dimensional:
$z$-component of the vector-potential is non-zero, hence another no-go result
prohibiting the 2+1 {\em gravitating} glueballs~\cite{deser} does not apply.

In the cylindrical case one can not ensure the conformal gauge for $g_{ab}$,
but one can always set $g_{\rho\rho}={\rm e}^\psi$
by a coordinate transformation. Denoting
\be
g_{zz}={\rm e}^\xi g_{\rho\rho},\quad \lambda= 2\phi-\psi,
\ee
we obtain from (\ref{Sax}) the following one-dimensional action:
\be
S=\int {\rm e}^{(\psi+\xi+\lambda)/2}\left(\psi'\xi'+\xi'\lambda'+\lambda'\psi'+
{\rm e}^{-\xi} R'^2 +{\rm e}^{-\lambda} P'^2  +  {\rm e}^{-(\xi+\lambda)}R^2P^2
\right)d\rho.
\ee
This corresponds to the  4D quantities
\bea
ds^2&=&-{\rm e}^{\psi} dt^2+d\rho^2+{\rm e}^{\lambda}d\varphi^2+
{\rm e}^{\xi}dz^2,\nonumber\\
A&=&R\;\tau_\varphi\, dz+ (P-\nu)\;\tau_z \,d\varphi,
\eea
where $\tau_\varphi$ is given by (\ref{taunu}) and primes denote derivatives
with respect to $\rho$.
Switching gravity off ($\psi=\xi=0,\,\lambda=\ln \rho^2$) leads exactly to the 
Nielsen-Olesen system with zero scalar self-coupling constant.

We conclude this section with a following remark. A cylindrical geometry does 
not prohibit the possibility of saddle points on the energy surface in the
configuration space. The reduced space dimensionality can be compensated 
in the minimax argument by  passing from non-contractible loops 
to non-contractible spheres. Such cylindrical sphalerons were found
indeed in the electroweak theory.
\section{Conclusions}
We have shown the stationary axisymmetric non-circular 4D SU(2) EYM system
admits a representation in terms of Euclidean 2D dilaton gravity coupled 
to a set of matter fields. The ansatz for the YM connection is specified
only by usual Kaluza-Klein toroidal assumptions suitably adapted to the 
non-Abelian case. New representation is aimed to facilitate further
consistent truncations of the model for different particular problems. 
As an application, a static cylindricaly symmetric ansatz is suggested 
as a candidate for a cylindrical EYM sphaleron. Our results may also be 
useful in the analysis of non-circular stationary axisymmetric spacetimes
with other matter sources.  
\section*{Acknowledgments}
The author would like to thank the Organizing Committee for an invitation
and  a stimulating atmosphere of the Workshop.  He also thanks the Yukawa
Institute for Theoretical Physics for hospitality while the final
version of the paper was written.  Fruitful discussions with
K.~Maeda, T.~Nakamura and M.~Sasaki and useful correspondence 
with M. Volkov are gratefully acknowledged.
The research was supported in part by the RFBR Grant 96--02--18899.

\end{document}